\documentclass[showpacs,pre,amsmath,amssymb]{revtex4}
\usepackage{latexsym}
\usepackage{amsmath}
\usepackage{amssymb}
\usepackage{graphicx}
\usepackage{dcolumn}
\usepackage{bm}
\graphicspath{{./}{./figs/}}

\begin{document}

\title{Synchronization of Coupled Nonidentical Genetic Oscillators\footnote{Physical Biology 3 (2006) 37-44.}}
\author{Chunguang Li$^{1,2,3}$}
\email{cgli@uestc.edu.cn }
\author{Luonan Chen$^4$}
\email{chen@elec.osaka-sandai.ac.jp}
\author{Kazuyuki Aihara$^{1,2}$}
\email{aihara@sat.t.u-tokyo.ac.jp} \affiliation{$^1$ERATO Aihara
Complexity Modelling Project, Room M204, Komaba Open Laboratory,
University of Tokyo,
4-6-1 Komaba, Meguro-ku, Tokyo 153-8505, Japan\\
$^2$Institute of Industrial Science, University of Tokyo, Tokyo 153-8505, Japan\\
$^3$Centre for Nonlinear and Complex Systems, University of
Electronic Science and Technology of China, Chengdu 610054, P. R.
China\\
$^4$Department of Electrical Engineering and Electronics, Osaka
Sanyo University, Osaka, Japan}
\date{\today}
\begin{abstract}
The study on the collective dynamics of synchronization among
genetic oscillators is essential for the understanding of the
rhythmic phenomena of living organisms at both molecular and
cellular levels. Genetic oscillators are biochemical networks,
which can generally be modelled as nonlinear dynamic systems. We
show in this paper that many genetic oscillators can be
transformed into Lur'e form by exploiting the special structure of
biological systems. By using control theory approach, we provide a
theoretical method for analyzing the synchronization of coupled
nonidentical genetic oscillators. Sufficient conditions for the
synchronization as well as the estimation of the bound of the
synchronization error are also obtained. To demonstrate the
effectiveness of our theoretical results, a population of genetic
oscillators based on the Goodwin model are adopted as numerical
examples.
\end{abstract}
 \maketitle
\section{INTRODUCTION}
Elucidating cooperative behavior of synchronization of coupled
genetic oscillators has important biological implications and
potential engineering applications from both theoretical and
experimental viewpoints, and it is also essential for the
understanding of the rhythmic phenomena of living organisms at
both molecular and cellular levels. So far many researchers have
studied the synchronization in genetic networks from the aspects
of experiment, numerical simulation and theoretical analysis. For
instance, in \cite{11}, the authors experimentally investigated
the synchronization of cellular clock in the suprachiasmatic
nucleus (SCN); in \cite{12} and \cite{13}, the collective dynamics
of synchronization are theoretically studied in synthetic
biological networks of identical genetic oscillators; and in
\cite{14}, the mechanism of synchronization in a population of
identical hysteretic genetic oscillators is analyzed.
Biologically, the genetic oscillators, even in the same species,
are usually nonidentical possibly due to asymmetrical nutrition
conditions and fluctuated environments, and the nonidentical
property can be modelled as parametric mismatches among
oscillators. For example, in the suprachiasmatic nucleus (SCN),
the periods of the circadian oscillators are not exactly the same,
and it has been observed that isolated individual neurons are able
to produce circadian oscillations with period ranging from 20 to
28 hours \cite{15, 16}. In \cite{8, 18}, the synchronization for
nonidentical genetic oscillators is examined numerically. Although
many mathematical models have been developed to study the
cooperative behaviors of cellular oscillation, there is no general
theoretical method in analyzing the dynamics of the coupled
genetic oscillators due to their inherent nonlinearity.

Genetic networks are biochemically dynamical systems, in which the
nodes indicate the biochemicals, and the couplings represent the
biochemical interactions \cite{1, 2}. Mathematically many genetic
oscillators can be expressed in the form of multiple additive
terms, each of which particularly is of linear, Michaelis-Menten
or Hill forms, such as the well-known Goodwin model \cite{3, 4},
repressilator \cite{5}, toggle switch \cite{6}, and the circadian
oscillators \cite{7}. From the synthetic biology viewpoint,
genetic oscillators with only linear, Michaelis-Menten and Hill
form terms can also be implemented easily. In this paper we
explore such special structure of gene networks to show that these
genetic oscillators can be transformed into Lur'e form and can be
further analyzed by using Lur'e system method in control theory
\cite{10}.

The aim of this paper is to provide a general theoretical method
for analyzing the synchronization of coupled nonidentical genetic
oscillators with the above-mentioned structure. In studying the
synchronization of genetic oscillators (and other nonlinear
systems), a general idea is to study the stability of the error
equations among oscillators. However, there are two main
difficulties for such method: (1) the difference of the oscillator
dynamics usually cannot be written into a function of the state
error; (2) due to the nonlinearity, there is no general efficient
analysis method for the stability of the error system. We show in
this paper that for genetic oscillators with the above-mentioned
structure, we can overcome both of the above difficulties. Since
coupled nonidentical oscillators usually cannot achieve complete
synchronization, a synchronous error is required to evaluate the
quality of the synchronization. In this paper, we present a
theoretical result, which can not only guarantee the
synchronization, but also estimates the bound of the
synchronization error. Besides, the obtained conditions can be
represented in terms of linear matrix inequalities (LMIs)
\cite{22}, which are very easy to be verified. Recently, it was
found that many biological networks are complex networks with
small-world and scale-free properties \cite{Tong, 20}. Our method
is also applicable to genetic oscillator networks with complex
topology, directed and weighted couplings. To demonstrate the
effectiveness of the theoretical results, we present two
simulation examples of coupled Goodwin oscillators with linear and
Michaelis-Menten couplings, respectively. Finally, several remarks
on the extensions of the proposed method are discussed. Notation
used in this paper as well as the detailed theoretical analysis
are given in the Appendix.

\section{METHODS AND RESULTS}
Mathematical modelling provides a powerful tool for studying gene
regulation processes in living organisms. Basically, there are two
types of genetic network models, i.e., the Boolean model (or
discrete model) and the differential equation model (or continuous
model) \cite{Smolen, DeJong, Bolouri}. In Boolean models, the
activity of each gene is expressed in one of two states, ON or
OFF, and the state of a gene is determined by a Boolean function
of the states of other related genes. In the differential equation
models, the variables describe the concentrations of gene
products, such as mRNAs and proteins, as continuous values, which
are more accurate and can provide detailed understanding of the
dynamical behaviors of the gene regulation systems. In this paper,
by adopting the differential equation models, we consider the
following form of a general genetic oscillator:
\begin{equation}
\dot{y}(t)=Ay(t)+\sum_{i=1}^l B_i f_i(y(t)),
\end{equation}
where $y(t)\in R^n$ represents the concentrations of proteins,
RNAs and chemical complexes, $A$ and $B_i$ are matrices in
$R^{n\times n}$, $f_i(y)=[f_{i1}(y_1(t)),\cdots,f_{in}(y_n(t))]^T$
with $f_{ij}(y_j(t))$ as a monotonic increasing or decreasing
regulatory function, which usually is of the Michaelis-Menten or
Hill form. Undoubtedly, many well-known genetic system models can
be represented in this form, such as the Goodwin model \cite{3,
4}, the repressilator \cite{5}, the toggle switch \cite{6}, and
the circadian oscillators \cite{7}. In synthetic biology, genetic
oscillators of this form can be implemented experimentally
\cite{Alon}. To make our method more understandable and to avoid
unnecessarily complicated notation, we consider the following
simplified model, in which there are only one increasing and one
decreasing nonlinear terms in each equation of the genetic
oscillator.
\begin{equation}
\dot{y}(t)=Ay(t)+B_1 f_1(y(t))+B_2 f_2(y(t)),
\end{equation}
where $Ay(t)$ includes the degradation terms and all the other
linear terms in the genetic oscillator, $f_1(y(t))=
[f_{11}(y_1(t)),\cdots,f_{1n}(y_n(t))]^T$ with $f_{1j}(y_j(t))$ as
a monotonic increasing function of the Hill form
\begin{equation*}
f_{1j}(y_j(t))=\frac{(y_j(t)/\beta_{1j})^{H_{1j}}}{1+(y_j(t)/\beta_{1j})^{H_{1j}}},
\end{equation*}
and $f_2(y(t))= [f_{21}(y_1(t)),\cdots,f_{2n}(y_n(t))]^T$ with
$f_{2j}(y_j(t))$ as a monotonic decreasing function of the
following form
\begin{equation*}
f_{2j}(y_j(t))=\frac{1}{1+(y_j(t)/\beta_{2j})^{H_{2j}}}.
\end{equation*}
In the above equations, both $H_{1j}$ and $H_{2j}$ are the Hill
coefficients. To avoid confusion, we let the $j$th column of
$B_{1,2}$ be zeros if $f_{1j,2j}\equiv 0$. Since
\begin{equation*}
f_{2j}(y_j(t))=\frac{1}{1+(y_j(t)/\beta_{2j})^{H_{2j}}}
=1-\frac{(y_j(t)/\beta_{2j})^{H_{2j}}}{1+(y_j(t)/\beta_{2j})^{H_{2j}}}\equiv1-g_{j}(y_j(t)),
\end{equation*}
by letting $f(.)=f_1(.)$, we can rewrite (2) as follows:
\begin{equation}
\dot{y}(t)=Ay(t)+B_1 f(y(t))-B_2 g(y(t))+B_2.
\end{equation}
Obviously, $f_i$ and $g_i$ satisfy the following sector
conditions
\begin{equation*}
\begin{array}{c}
0\leq \frac{f_i(a)-f_i(b)}{a-b} \leq k_{1i},\\
0\leq \frac{g_i(a)-g_i(b)}{a-b} \leq k_{2i},\\
\forall a, b \in R\, (a\neq b); i=1,\cdots,n,
\end{array}
\end{equation*}
or equivalently,
\begin{equation}
\begin{array}{c}
(f_i(a)-f_i(b))[(f_i(a)-f_i(b))-k_{1i}(a-b)] \leq 0,\\
(g_i(a)-g_i(b))[(g_i(a)-g_i(b))-k_{2i}(a-b)] \leq 0,\\
\forall a, b\in R\, (a\neq b); i=1,\cdots,n,
\end{array}
\end{equation}
It follows from the mean value theorem that for differentiable
$f_i$ and $g_i$, the above sector conditions correspond to
\begin{equation}
\begin{array}{c}
0\leq \frac{df_i}{da}(a) \leq k_{1i},\\
0\leq \frac{dg_i}{da}(a)\leq k_{2i},\\
\forall a \in R; i=1,\cdots,n.
\end{array}
\end{equation}

Recall that a Lur'e system is a linear dynamic system, feedback
interconnected to a static nonlinearity $f(.)$ that satisfies a
sector condition \cite{10}. Hence, the genetic oscillator (3) can
be seen as a Lur'e system, which can be investigated by using the
fruitful Lur'e system method in control theory. In the following,
we first consider coupled identical genetic oscillators, and then
extend the result to the nonidentical case.

We first analyze the following $N$ linearly coupled genetic
oscillators, in which each genetic oscillator is identical.
\begin{equation}
\dot{x}_i(t)=Ax_i(t)+B_1 f(x_i(t))-B_2
g(x_i(t))+B_2+\sum_{j=1}^NG_{ij}Dx_j(t),\, i=1,\cdots,N
\end{equation}
where $x_i(t)\in R^n$ is the state vector of the $i$th genetic
oscillator (corresponds to $y(t)$ in Eq. (3)), $D\in R^{n\times
n}$ defines the coupling between two genetic oscillators.
$G=(G_{ij})_{N\times N}$ is the coupling matrix of the network, in
which $G_{ij}$ is defined as follows: if there is a link from
oscillator $j$ to oscillator $i$ $(j\neq i)$, then $G_{ij}$ equals
to a positive constant denoting the coupling strength of this
link; otherwise, $G_{ij}=0$; $G_{ii}=-\sum_{j=1}^n G_{ij}$. Matrix
$G$ defines the coupling topology, direction, and the coupling
strength of the network.

Since in biological networks, the genetic oscillators are usually
nonidentical, there are parametric mismatches among oscillators.
Next, we consider the following network of $N$ coupled
nonidentical genetic oscillators:
\begin{equation}\begin{array}{c}
\dot{x}_i(t)=(A+\Delta A_i(t))x_i(t)+(B_1+\Delta B_{1i}(t))
f(x_i(t))-(B_2+\Delta B_{2i}(t)) g(x_i(t))+(B_2+\Delta B_2(t))\\
+\sum_{j=1}^NG_{ij}Dx_j(t),\, i=1,\cdots,N
\end{array}
\end{equation}
where $\Delta A_i, \Delta B_{1i}, \Delta B_{2i}$ are the mismatch
matrices, which can be time-varying. We assume that the mismatch
matrices $\Delta A_i(t), \Delta B_{1i}(t), \Delta B_{2i}(t)$ can
be estimated by the following bounds, which are also reasonable
for general biological systems.
\begin{equation*}
\|\Delta A_i(t)\|\leq \alpha_1, \|\Delta B_{1i}(t)\|\leq
\alpha_2,\|\Delta B_{2i}(t)\|\leq \alpha_3, \forall i .
\end{equation*}
We also assume that
\begin{equation*}
\|x_i(t)\|\leq \delta_1, \|f(x_i(t))\|\leq \delta_2,
\|g(x_i(t))\|\leq \delta_3, \forall i .
\end{equation*}
Since in genetic oscillators, $x_i(t)$ usually denotes the
concentrations of {\it m}RNA, protein, neurotransmitter, etc.,
which are of limited values, and $f(.)$ and $g(.)$ are usually
monotonic functions with saturated values, the above assumptions
are also reasonable. The other parameters are defined as the same as those in the
identical case.

For the above two network, i.e. Eq. (6) and (7),  we mainly use
Wu's method to analyze the synchronization \cite{19}, which can
separate the effects of the coupling and the individual genetic
oscillator dynamics. Based on Lyapunov method, we can obtain the
following sufficient conditions for the synchronization of coupled
identical genetic oscillators (Eq. (6)) and coupled nonidentical
genetic oscillators (Eq. (7)) in Theorems 1 and 2, respectively.
In the following theorems and hereafter,
$K_1=\mbox{diag}(k_{11},\cdots,k_{1n})$,
$K_2=\mbox{diag}(k_{21},\cdots,k_{2n})$, and matrix $U\in
R^{N\times N}$ is defined as an irreducible matrices with zero row
sums, whose off-diagonal elements are all non-positive.
$\lambda_{min}(P)$ and $\lambda_{max}(P)$ represent the minimal
and maximal eigenvalues of the matrix $P$ respectively, and
$\otimes$ indicates the  Kronecker product, which is defined in
Appendix.

\emph{Theorem 1}: If there are matrices  $P>0$,
$\Lambda_1=\mbox{diag}(\lambda_{11},\cdots,\lambda_{1n})>0$,
$\Lambda_2=\mbox{diag}(\lambda_{21},\cdots,\lambda_{2n})>0$, $Q\in
R^{n\times n}$, and $U\in R^{N\times N}$, such that the following
matrix inequalities hold
\begin{equation}
\begin{array}{l}
M_1=\left[\begin{array}{ccc}PA+A^TP-Q-Q^T &PB_1+K_1\Lambda_1
&-PB_2+K_2\Lambda_2\\B_1^TP+K_1\Lambda_1&-2\Lambda_1&0 \\
-B_2^TP+K_2\Lambda_2&0&-2\Lambda_2\end{array}\right]<0,\\
(UG\otimes PD+U\otimes Q)^T+(UG\otimes PD+U\otimes Q)\leq 0,
\end{array}
\end{equation}
then the network (6) is asymptotically synchronous.

\emph{Theorem 2}: If there are matrices  $P>0$,
$\Lambda_1=\mbox{diag}(\lambda_{11},\cdots,\lambda_{1n})>0$,
$\Lambda_2=\mbox{diag}(\lambda_{21},\cdots,\lambda_{2n})>0$, $Q\in
R^{n\times n}$, $U\in R^{N\times N}$, and a positive real constant
$\gamma$ such that the following matrix inequalities hold
\begin{equation}
\begin{array}{l}
M_2=\left[\begin{array}{ccc}PA+A^TP-Q-Q^T+\gamma I
&PB_1+K_1\Lambda_1
&-PB_2+K_2\Lambda_2\\B_1^TP+K_1\Lambda_1&-2\Lambda_1&0 \\
-B_2^TP+K_2\Lambda_2&0&-2\Lambda_2\end{array}\right]<0,\\
(UG\otimes PD+U\otimes Q)^T+(UG\otimes PD+U\otimes Q)\leq 0,
\end{array}
\end{equation}
then the network (7) is asymptotically synchronous with error
bound
\begin{equation*}
\sum_{i<j}(-U_{ij})\|x_i(t)-x_j(t)\|^2\leq \frac{\beta ^2}{\gamma
^2}\frac{\lambda_{max}(P)}{\lambda_{min}(P)}\sum_{i<j}(-U_{ij})
\end{equation*}
where
\begin{equation*}
\beta=4(\alpha_1\delta_1+\alpha_2\delta_2+\alpha_3\delta_3+\alpha_3)\lambda_{max}(P)
\end{equation*}

The proof of the above theorems are somewhat technical, and we
defer the details to the Appendix. In Theorem 2, we not only give
a sufficient condition for the synchronization, but also provide
an estimation of the synchronization error bound. In this error
bound estimation, we can select the form of the matrix $U$
beforehand to obtain different error combinations. Specifically,
if we choose the following form of $U$
\begin{equation*}
U=\left[
\begin{array}{cccc}
N-1&-1&\cdots&-1\\
-1&1& &\\
\vdots& &\ddots& \\
-1& & &1\end{array} \right]
\end{equation*}
we can obtain the following synchronous error estimation:
\begin{equation*}
\sum_{j=2}^N\|x_j(t)-x_1(t)\|^2\leq
(N-1)\frac{\beta^2}{\gamma^2}\frac{\lambda_{max}(P)}{\lambda_{min}(P)}.
\end{equation*}

It should be note that, since in each step, we used conservative
estimations of the bounds, the estimated error bound may be much
larger than the actual error. In other words, if accurate
information on the parametric mismatches is known, we can have a
better error estimation.

The first matrix inequality in (8) (or (9)) is an LMI, which can
be easily verified by using convex optimization techniques, e.g.,
the interior point method \cite{22}, and by software packages,
e.g., the MATLAB LMI Toolbox. If we choose the matrix $U$
beforehand, the second matrix inequality in (8) (or (9)) is also
an LMI. Furthermore, for two special cases, we have the following
results \cite{23}:

(1) When $G$ is symmetric: Letting $U=-G$, the second matrix
inequality in (8) (or (9)) can be rewritten as
\begin{equation*}
-G^2\otimes (PD+D^TP)-G\otimes (Q+Q^T)\leq 0,
\end{equation*}
which is equivalent to
\begin{equation*}
\sigma_i(PD+D^TP)+(Q+Q^T)\leq 0, \mbox{for all nonzero eigenvalues
} \sigma_i \mbox{ of } G.
\end{equation*}

(2) When $D$ is symmetric and commutable with $G$: Letting
$U=-(G+G^T)/2$, the second matrix inequality in (8) (or (9)) can
be rewritten as
\begin{equation*}
-\frac{1}{2}(G+G^T)^2\otimes PD-\frac{1}{2}(G+G^T)\otimes
(Q+Q^T)\leq 0,
\end{equation*}
which is equivalent to
\begin{equation*}
2\sigma_iPD+(Q+Q^T)\leq 0, \mbox{for all nonzero real part }
\sigma_i \mbox{ of the eigenvalues of } G.
\end{equation*}

Thus the second matrix inequality in (8) (or (9)) is also a lower
dimensional LMI, which together with the first LMI can be verified
easily.

Next, we use numerical examples to show the effectiveness of the
theoretical results.

\section{NUMERICAL EXAMPLES}
To demonstrate the effectiveness of our theoretical methods, we
study a population of coupled SCN neuron model oscillators. The
single cell or genetic oscillator is described by the classical
Goodwin model \cite{3}. In this model, a clock gene mRNA ($X$)
produces a clock protein ($Y$), which activates a transcriptional
inhibitor ($Z$). $Z$ inhibits the transcription of the clock gene,
thus forming a negative feedback loop. In this paper, we assume
that the light $L=0$. The oscillators coupled through the release
and receiving of neurotransmitter among neurons. Similar to
\cite{18}, the evolution equations for a network of $N$ coupled
nonidentical oscillators are given below:
\begin{equation}
\begin{array}{l}
\dot{X}_i=v_{i1}\frac{1}{1+Z_i^m}-v_{i2}X_i+KF,\\
\dot{Y}_i=v_{i3}X_i-v_{i4}Y_i,\\
\dot{Z}_i=v_{i5}Y_i-v_{i6}Z_i,\\
\dot{V}_i=v_{i7}X_i-v_{i8}V_i,
\end{array}
\end{equation}
where $v_{i1},v_{i2},v_{i3},v_{i4},v_{i5},v_{i6},v_{i7},v_{i8}$
are positive constants, and $K>0$ is the coupling strength. The
variables $X_i, Y_i, Z_i$ describe the dynamics of the oscillator
in the $i$th neuron, and $V_i$ describes the evolution of the
neurotransmitter in the $i$th neuron. The release of the
neurotransmitter is supposed to be fast with respect to the 24-h
timescale of the oscillators and becomes homogeneous to establish
an average neurotransmitter level, or a mean field $F$ \cite{18}
\begin{equation*}
F=\frac{1}{N}\sum_{j=1}^NV_j.
\end{equation*}

Clearly, the individual Goodwin model is of the form (3), in which
$f\equiv 0$, $B_1=0$, $g=[0,0,Z_i^m/(1+Z_i^m),0]^T$, $B_2$ is a $4
\times 4$ matrix with all zero entries except for
$B_2(1,3)=v_{i1}$, and all the other terms are in the linear form.
By plus and minus $KV_i$ in the first equation of (10), we can get
the coupling term $\frac{K}{N}\sum_{j=1}^N(V_j-V_i)$.

The purpose of this section is to demonstrate the effectiveness of
the theoretical method, instead of mimicking the real SCNs. We
consider a small size of network with $N=10$ coupled oscillators,
although there are $\sim 10^4$ neurons in the SCNs. The
concentrations are expressed in $nM$, and the standard parameters
are set as $m=12,v_{i1}=1\, nM/h,v_{i3}=v_{i5}=1/h,
v_{i7}=0.2/h,v_{i2}=0.3/h,v_{i4}=0.22/h,v_{i6}=0.15/h,v_{i8}=2/h$
for all $i$, so that the period of the oscillator is approximately
24h, and the coupling strength is $K=1$. It is known that the
period of the Goodwin model is sensitive to the parameters
$v_{i2},v_{i4},v_{i6}$ \cite{4}. The mismatches of
$v_{i2},v_{i4},v_{i6}$ are randomly distributed in $\pm 10\%$
around the above values of $v_{i2},v_{i4},v_{i6}$. In Fig. 1, when
starting from the same initial values, we show the oscillation
dynamics of the mRNA concentrations of the 10 uncoupled
oscillators, which indicate that the periods of the oscillators
are quite different. Submitting the above parameters to the
corresponding matrices in the matrix inequalities (9) of Theorem
2, letting $U=-G$ and using MATLAB LMI Toolbox, we can easily find
feasible solutions for (9), which indicate that the above
all-to-all coupled network can achieve synchronization although it
is not a complete synchronous state. In Fig. 2 (a), when starting
from different initial values, we plot the time evolution of the
mRNA concentrations ($X_i$) of all the oscillators. Fig. 2 (b)
shows the synchronization error
\begin{equation*}
J=\sum_{i=2}^N[(X_i-X_1)^2+(Y_i-Y_1)^2+(Z_i-Z_1)^2+(V_i-V_1)^2].
\end{equation*}
which is gradually reduced with time evolution.

\begin{figure*}[htb]
\centering
\includegraphics[width=16cm]{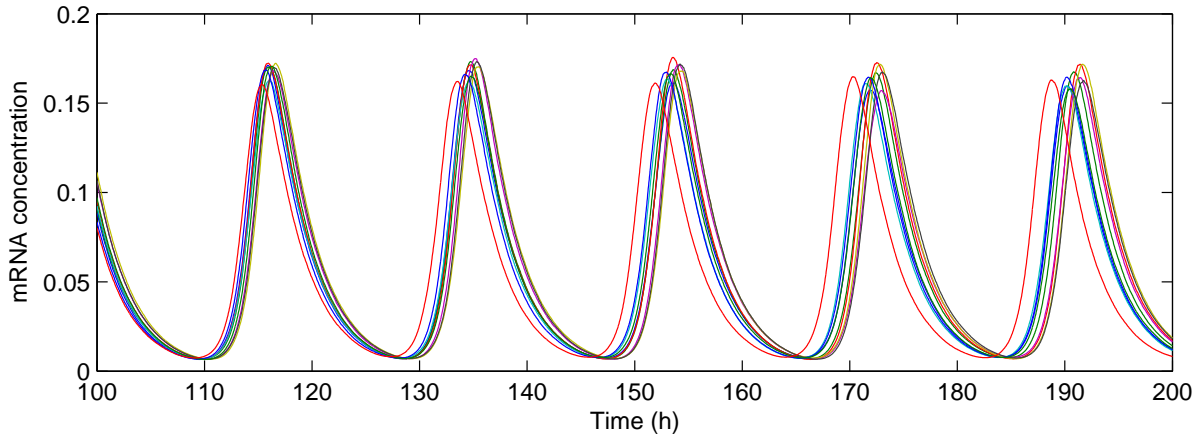}
\caption{Oscillation dynamics of the mRNA concentrations of the
uncoupled oscillators with the same initial conditions.}
\end{figure*}

\begin{figure*}[htb]
\centering
\includegraphics[width=16cm]{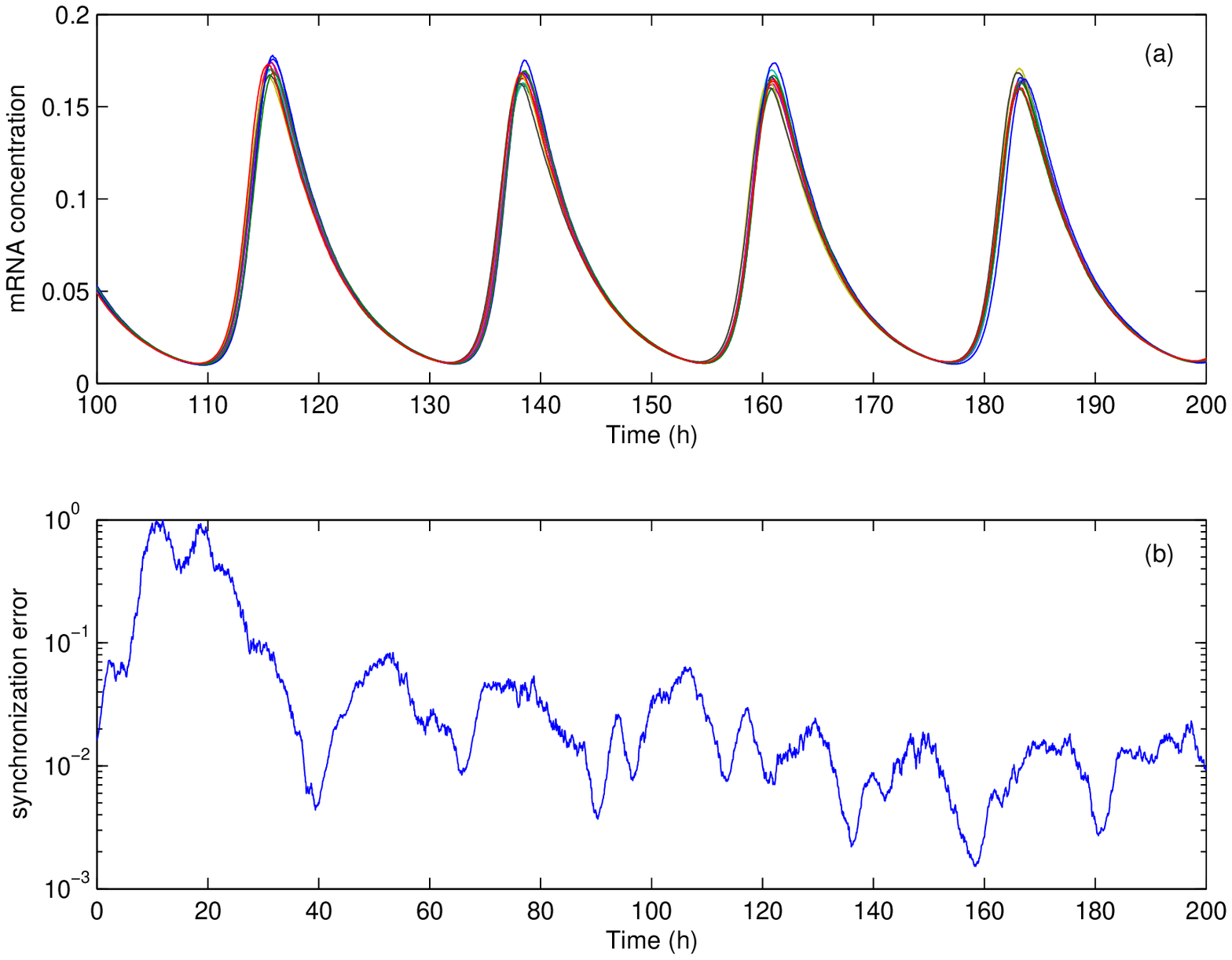}
\caption{Time evolution of Goodwin models with linear coupling.
(a)The time evolution of the mRNA concentrations of the
oscillators; (b) The time evolution of the synchronization error.}
\end{figure*}

\begin{figure*}[htb]
\centering
\includegraphics[width=16cm]{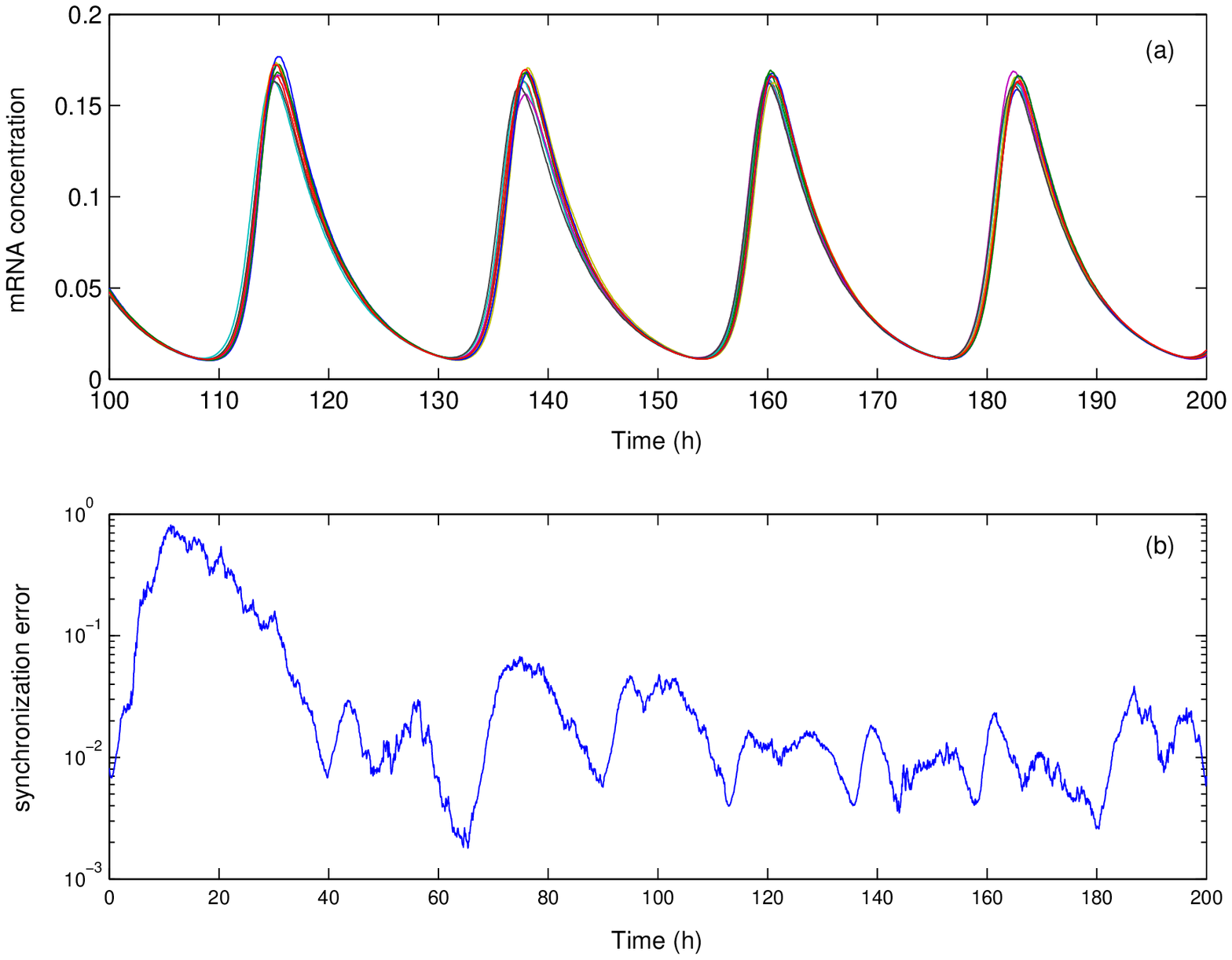}
\caption{Time evolution of Goodwin models with Michaelis-Menten
coupling. (a)The time evolution of the mRNA concentrations of the
oscillators; (b) The time evolution of the synchronization error.}
\end{figure*}

Since there is a maximal activity of fully active promoters, in
\cite{18}, the authors considered a Michaelis-Menten form of
coupling term, that is, in (10) replacing $KF$ by
$K\frac{F}{1+F}$. Our theoretical results in Theorems 1 and 2 are
also applicable for this case. Specifically, we may have different
methods to treat the coupling term. One of the simplest ways is
explained as follows: Since the coupling term $K\frac{F}{1+F}$ is
the same for all oscillators and $U$ is zero row sums, it is easy
to show that the product of $U\otimes P$ and the vector of the
coupling terms (containing the coupling terms of all the
oscillators) is zero. By using the same analysis as that of
Theorem 2, it is easy to show that the synchronization condition
is only the first LMI in Theorem 2 with $Q=0$. For the
Michaelis-Menten coupling, Figs. 3 (a) and 3 (b) show the time
evolution of the $m$RNA concentrations and the synchronization
error, respectively. Figs. 2 and 3 indicate that the coupled
oscillators are indeed synchronized with small error bounds, which
confirms the theoretical results.

\section{CONCLUSION AND OUTLOOK}
In this paper, we presented a theoretical method for analyzing the
synchronization of coupled nonidentical genetic oscillators based
on control theory approach. The purpose of this paper is not to
provide a general theory for all genetic oscillator networks, but
provide an efficient method for genetic oscillator networks that
can be expressed in the form of (1). In addition, the sufficient
conditions for the synchronization were also derived based on LMI
formalism, which can be easily verified numerically. Although the
method is proposed for genetic oscillator networks, it is also
applicable to other biochemical and neuronal networks of the form.
To make the theoretical method more understandable and to avoid
unnecessarily complicated notation, we discussed only on some
simplified forms of the genetic oscillators, but more general
cases and extensions regarding this topic can be studied in a
similar way, for example:
\begin{enumerate}
\item  The theoretical results can be easily extended to the
general form of genetic oscillators with more than 2 nonlinear
terms in each equation as shown in (1).

\item The genetic oscillator model (3) can be extended to a more
general case such that $f_{i}$ and $g_{i}$, the components of
$f(y(t))$ and $g(y(t))$, are functions of $y(t)$, instead of
$y_i(t)$. For this case, we only require that
\begin{equation*}
\begin{array}{c}
0\leq \frac{f_i(a)-f_i(b)}{c_{1i}^T(a-b)} \leq k_{1i}\\
0\leq \frac{g_i(a)-g_i(b)}{c_{2i}^T(a-b)} \leq k_{2i}\\
\forall a, b \in R^n\, (a\neq b); i=1,\cdots,n
\end{array}
\end{equation*}
where $c_{1i},c_{2i}\in R^n$ are arbitrary nonzero real vectors.
Moreover, $f_{i}$ and $g_{i}$ can be of more complex forms
(non-Hill form) and non-differentiable provided that they satisfy
sector conditions.

\item Genetic oscillators who have a few terms that are not in
linear, Michaelis-Menten and Hill forms, can also be analyzed
similarly by using our method. For example, in the mammalian
circadian clock model \cite{24}, there are a few product terms (of
the form $xy$) besides the three kinds of terms. We can treat
these terms as follows: Since
\begin{equation*}
x_1y_1-x_2y_2=x_1(y_1-y_2)+y_2(x_1-x_2),\mbox{ and }
\frac{x_1(y_1-y_2)}{y_1-y_2}\leq max(x_1),
\frac{y_2(x_1-x_2)}{x_1-x_2}\leq max(y_2),
\end{equation*}
then we can treat these terms similarly as those that satisfies
the sector condition. Although such manipulation introduces
additional conservation, the conservation is assumed to be limited
since most terms in the model are of the three kind of terms.
\end{enumerate}

In addition to the nonlinear and coupling properties, genetic
networks are intrinsically noisy \cite{25,26,27,27a,28}, and with
significant time delays \cite{2a,4a}. Future works regarding this
topic also include the extension of our method to the case with
noise perturbations and time delays.

\section*{ACKNOWLEDGEMENT}
This research was supported by Grant-in-Aid for Scientific
Research on Priority Areas 17022012 from MEXT of Japan, the
National Natural Science Foundation of China under Grant 60502009,
and the Program for New Century Excellent Talents in University of
China.

\section*{GLOSSARY}

Gene regulatory network or genetic network: A gene regulatory
network is a collection of DNA segments in a cell which interact
with each other and with other substances in the cell, thereby
governing the rates at which genes in the network are transcribed
into mRNA.

Circadian rhythm: Circadian rhythm is the name given to the
roughly 24 hour cycles shown by physiological processes in plants,
animals, fungi and cyanobacteria.

Synchronization: synchronization of dynamical systems refers to a
process wherein two (or many) systems (either identical or
nonidentical) adjust a given property of their motion to a common
behavior due to a coupling or to a forcing (periodical or noisy).

Lur'e system: a Lur'e system is a linear dynamic system, feedback
interconnected to a static nonlinearity that satisfies a sector
condition.

\section*{APPENDIX}

\subsection*{A. Notations:}
Throughout this paper, $A^T$ denotes the transpose of a square
matrix $A$. The notation $M > (<)\,0$ is used to define a real
symmetric positive definite (negative definite) matrix. $R^m$
denotes the $m$-dimensional Euclidean space; and $R^{n\times m}$
denotes the set of all $n\times m$ real matrices. In this paper,
if not explicitly stated, matrices are assumed to have compatible
dimensions. $\|\cdot\|$ stands for the usual $L_2$ norm of a
vector, or the usual spectral norm of a square matrix. Matrix
$U\in R^{N\times N}$ is defined as an irreducible matrices with
zeros row sum, whose off-diagonal elements are all non-positive.
The Kronecker product $ A \otimes B$ of an $n \times m$ matrix $A$
and a $p \times q$ matrix $B$ is the $np \times mq$ matrix defined
as
\begin{equation*}
A\otimes B=\left[\begin{array}{ccc} A_{11}B& \cdots &A_{1m}B\\
\vdots &\ddots &\vdots\\ A_{n1}B &\cdots &
A_{nm}B\end{array}\right]
\end{equation*}

\subsection*{B. Theoretical Analysis of the Synchronization}

\emph{Proof of Theorem 1}: We let
$x(t)=[x_1^T(t),\cdots,x_N^T(t)]^T \in R^{Nn\times 1}$, and define
a Lyapunov function of the following form:
\begin{equation}
V(x(t))=x^T(t)(U\otimes P)x(t)
\end{equation}

According to \cite{19} (pp.136-137), $V(x(t))$ is equivalent to
the following form
\begin{equation*}
V(x(t))=\sum_{i<j}(-U_{ij})(x_i(t)-x_j(t))^T P(x_i(t)-x_j(t)).
\end{equation*}
Hence, if the time derivative of $V(x(t))$ along the trajectories of
 (6) is negative, then according to Lyapunov's direct
method, the genetic oscillators will achieve synchronization.

Calculating the time derivative of $V(x(t))$, we have
\begin{equation}
\begin{array}{rl}
\dot{V}(x(t))=&2\sum_{i<j}(-U_{ij})(x_i(t)-x_j(t))^T
P[(A-T)(x_i(t)-x_j(t))+B_1(f(x_i)-f(x_j))-B_2(g(x_i)-g(x_j))]\\
&+2x^T(t)(U\otimes P)(G\otimes D+I\otimes T)x(t)
\end{array}
\end{equation}
where $T\in R^{n\times n}$ is an arbitrary real matrix. For all
$i,j\,(i\neq j)$, we have
\begin{equation*}
\begin{array}{l}
L_{ij}=2(x_i(t)-x_j(t))^T
P[(A-T)(x_i(t)-x_j(t))+B_1(f(x_i(t))-f(x_j(t)))-B_2(g(x_i(t))-g(x_j(t)))]\\
\leq 2(x_i(t)-x_j(t))^T
P(A-T)(x_i(t)-x_j(t))+2(x_i(t)-x_j(t))^TPB_1(f(x_i(t))-f(x_j(t)))\\
\hspace{0.3cm} -2(x_i(t)-x_j(t))^TPB_2(g(x_i(t))-g(x_j(t)))\\
\hspace{0.3cm}-2\sum_{l=1}^n\lambda_{1l}(f_l(x_{il}(t))-f_l(x_{jl}(t)))[(f_l(x_{il}(t))-f_l(x_{jl}(t)))-k_{1l}(x_{il}(t)-x_{jl}(t))]\\
\hspace{0.3cm}-2\sum_{l=1}^n\lambda_{2l}(g_l(x_{il}(t))-g_l(x_{jl}(t)))[(g_l(x_{il}(t))-g_l(x_{jl}(t)))-k_{2l}(x_{il}(t)-x_{jl}(t))]\\
\end{array}
\end{equation*}
By letting $Q=PT$, we have
\begin{equation*}
L_{ij}=\xi_{ij}(t)M_1\xi_{ij}(t)<0
\end{equation*}
for all $i,j$.
$\xi_{ij}=[(x_i(t)-x_j(t))^T,(f(x_i(t))-f(x_j(t)))^T,
(g(x_i(t))-g(x_j(t)))^T]^T\in R^{3n\times 1}$. Therefore, the
first term of (12) is negative except for $x_i(t)=x_j(t)$,
$\forall i,j$. By letting $Q=PT$, $(U\otimes P)(G\otimes
D+I\otimes T)^T+(U\otimes P)(G\otimes D+I\otimes T)$ is equivalent
to the second matrix inequality in (8), which means that the
second term of (12) is non-positive. We have $\dot{V}(x(t))<0$ in
(12). Thus the Theorem 1 is proved. $\Box$

Next, we consider the case of coupled nonidentical genetic
oscillators (7). In \cite{21}, the authors studied the robust
synchronization of master-slave coupled two nonidentical chaotic
systems of the Lur'e form. Here we extend the result to the case
of complex networks, and apply it to the coupled nonidentical
genetic oscillators.

{\it Proof of Theorem 2}: We also use the Lyapunov function (11).
By calculating the time derivative of $V(x(t))$, we have
\begin{equation}
\begin{array}{rl}
\dot{V}(x(t))=&2\sum_{i<j}(-U_{ij})(x_i(t)-x_j(t))^T P[((A+\Delta
A_i(t))-T)(x_i(t)-x_j(t))\\&\hspace{1cm}+(B_1+\Delta
B_{1i}(t))(f(x_i)-f(x_j))-(B_2+\Delta B_{2i}(t))(g(x_i)-g(x_j))+(\Delta B_{2i}-\Delta B_{2j})]\\
&+2x^T(t)(U\otimes P)(G\otimes D+I\otimes T)x(t)
\end{array}
\end{equation}
By letting $Q=PT$, it is easy to know the second term is
nonpositive, and for all $i,j\,(i\neq j)$, we have
\begin{equation*}
\begin{array}{l}
2(x_i(t)-x_j(t))^T P[((A+\Delta
A_i(t))-T)(x_i(t)-x_j(t))+(B_1+\Delta
B_{1i}(t))(f(x_i)-f(x_j))\\\hspace{3cm}-(B_2+\Delta
B_{2i}(t))(g(x_i)-g(x_j))+(\Delta B_{2i}-\Delta B_{2j})]\\
=2(x_i(t)-x_j(t))^T
P[(A-T)(x_i(t)-x_j(t))+B_1(f(x_i(t))-f(x_j(t)))-B_2(g(x_i(t))-g(x_j(t)))]\\
\hspace{0.3cm}+2(x_i(t)-x_j(t))^T P [\Delta A_i(t)x_i(t)-\Delta
A_j(t)x_j(t)+\Delta B_{1i}(t)f(x_i(t))-\Delta
B_{1j}(t)f(x_j(t))\\
\hspace{0.3cm}-\Delta B_{2i}(t)g(x_i(t))+\Delta
B_{2j}(t)g(x_j(t))+\Delta B_{2i}(t)-\Delta B_{2j}(t)]\\
\leq
\xi_{ij}(t)M_1\xi_{ij}(t)+2\|x_i(t)-x_j(t)\|\lambda_{max}(P)(2\alpha_1\delta_1+2\alpha_2\delta_2+2\alpha_3\delta_3+2\alpha_3)\\
=\xi_{ij}(t)M_2\xi_{ij}(t)-\gamma\|x_i(t)-x_j(t)\|^2+\beta\|x_i(t)-x_j(t)\|\\
<-\gamma\|x_i(t)-x_j(t)\|^2+\beta\|x_i(t)-x_j(t)\|
\end{array}
\end{equation*}
where $\xi_{ij}(t)$ and $M_1$ are the same as those defined in the
Proof of Theorem 1. We have $\dot{V}(x(t))<0$ if
$\|x_i(t)-x_j(t)\|\geq \beta/\gamma$. Since
\begin{equation*}
\lambda_{min}(P)\sum_{i<j}(-U_{ij})\|(x_i(t)-x_j(t))\|^2\leq
V(x(t))\leq
\lambda_{max}(P)\sum_{i<j}(-U_{ij})\|(x_i(t)-x_j(t))\|^2,
\end{equation*}
we have
\begin{equation*}
\lambda_{min}(P)\sum_{i<j}(-U_{ij})\|(x_i(t)-x_j(t))\|^2\leq
\lambda_{max}(P)\frac{\beta^2}{\gamma^2}\sum_{i<j}(-U_{ij}).
\end{equation*}
Thus we obtain the error bound in Theorem 2. Note that
$\frac{\lambda_{max}(P)}{\lambda_{min}(P)}$ is the conditional
number of matrix $P$. $\Box$

\end{document}